\newcommand{\BF}[1]{\mbox{\boldmath $#1$}}

\documentstyle[eqsecnum,aps]{revtex}

\def\ins#1{}

\def\iems#1{}
\def\comment#1{}

\def\cm#1{}
\def\sbf#1{\mbox{\scriptsize{\bf #1}}}

\begin{document}
\setcounter{figure}{0}
\Roman{figure}

\title{{
Quantum-Liouville and Langevin Equations for Gravitational Radiation Damping
}}
\author{Z. Haba%
 \thanks{On leave from Institute of Theoretical Physics, University of
Wroclaw, Poland; e-mail: zhab@ift.uni.wroc.pl}
 and H. Kleinert%
 \thanks{Email: kleinert@physik.fu-berlin.de 
URL:
http://www.physik.fu-berlin.de/\~{}kleinert \hfill
} }
\address{Institut f\"ur Theoretische Physik,\\
Freie Universit\"at Berlin, Arnimallee 14,
14195 Berlin, Germany}
\maketitle
\begin{abstract}
{}From  a forward--backward path integral, we derive a master equation for the
emission and
absorption of gravitons
by a massive quantum object in a heat
bath of gravitons. Such an equation could describe collapse phenomena of dense
stars.
We also present a
useful approximate Langevin equation for
such a system.
\end{abstract}

%
\section{Introduction}
Usually, quantum effects are
negligible for objects of a large mass which consist
of a great number of particles. Under special conditions, however,
such objects can
be of microscopic size.
Then the weak gravitational forces may
 sum up coherently.
The process of a collapse
of a star will be accompanied by an
emission of gravitational waves which should be treated quantum
mechanically in the final stages, e.g.~during the formation
of primordial black holes.
In this paper we set up an equation  for describing
the behavior of a gravitational
object of this kind when interacting
with an equilibrium state of gravitons.
\section{Path integral for density matrix}
Consider the Schr\"odinger equation
\begin{eqnarray}
 i\hbar \partial_t
\langle {\bf x}|\psi (t) \rangle=  \hat  H(t) \langle {\bf x}|\psi (t) \rangle,
\label{2.1}\end{eqnarray}
where
$
 \hat H(t) = -\hbar ^2   \Delta /2M   + V({\bf x})
$ 
 is the Hamilton operator
formed from the Laplace-Beltrami operator
$ \Delta =g^{-1/2} \left( \partial_\mu
  g^{\mu \nu }  {g}^{1/2}  \partial_ \nu \right)$
containing the
inverse $g^{\mu \nu }$ of the metric  $g_{\mu \nu }({\bf x},t)$
 and the determinant $g=$\,det$ \,(g_{\mu \nu })$.
In empty space, small fluctuations $2 \varepsilon \,u_{\mu \nu }(x)$ of the
metric around
the
Minkowski metric $ \eta _{\mu \nu }$
describe gravitational
waves
which may be quantized as usual.
In the linear approximation,
the Einstein action
reads
\begin{eqnarray}
   {\cal A}_{\rm E}  =  -\frac{1}{2  \varepsilon ^2 } \int d^4x
         \sqrt{ -g} R
=  - \frac{1}{2  }
	\int d^4x\, G^{\mu \nu }u_{\mu \nu },~~~~G^{\mu \nu }\equiv
\epsilon _{\mu  \lambda  \sigma \tau }
\epsilon _{ \nu  \lambda  \kappa  \delta }\partial _ \sigma  \partial_ \kappa
 u_{\tau  \delta },
\label{lingr5.13}\end{eqnarray}
  where $
G^{\mu \nu }$ is the linearized
Einstein tensor $R^{\mu \nu }-g^{\mu \nu }R/2$, and
$\varepsilon$
 a gravitational
coupling constant related to Newton's gravitational constant
  $  \varepsilon ^2=8\pi G/c^3$.
In the radiation gauge with $u^{0a}\equiv 0$, the
free fields
may be expanded into creation and annihilation operators
for gravitons
as
\begin{equation}
\hat u_{\mu\nu}(x)=
\sum _{{\bf k}}
\sum _{h=-2,2}
\frac{1}{ \sqrt{2V \Omega _{\sbf k}} }
\left[
 \epsilon _{\mu\nu}(\hat{\bf k},h)
\hat a(\hat{\bf k},h)e^{-ikx}+{\rm h.c.}
\right] ,
\label{@}\end{equation}
where
$
 \epsilon _{\mu\nu}(\hat{\bf k},\pm2)$ are the
traceless divergence-free polarization tensors of helicity $\pm2$.
The
correlation functions at temperature $T$ of these fields are:
\begin{eqnarray}
\! \langle\hat T \hat u_{ij} ({\bf x},t) \hat u_{kl} ({\bf x}', t')\rangle
\equiv G_{ij;kl} ({ x}, { x}') & \!=\! &
  c {\hbar }                            \!\!
     \int\! \!\frac{d^3k}{(2\pi)^3 \Omega _{\sbf k}} \Lambda_{ij;kl}
(\hat {\bf k})
  \cos  {\bf k}
    ( {\bf x} \!-\!{\bf x}')  \!
   \left[ \cos  \Omega _{\sbf k}  (t \!-\! t')
       \coth  \frac{\beta\hbar  \Omega _{\sbf k}}{2}
 \!-\! i \sin \Omega _{\sbf k}( t\!-\!t')\right],
\nonumber \\
\label{2.4}\end{eqnarray}
where    $ \Omega _k = c|{\bf k}|$ and
\begin{eqnarray}
 \Lambda_{ij,kl}(\hat{{\bf k}})&\equiv &
 \sum _{h=-2,2}\epsilon(\hat {\bf k},h) \epsilon(\hat {\bf k},h)^*=
  \frac{1}{2}
\left[\Lambda_{ik} (\hat {\bf k}) \Lambda_{jl}  (\hat {\bf k})+
  \Lambda_{il}  (\hat {\bf k}) \Lambda_{jk}  (\hat {\bf k})
   - \Lambda_{ij}  (\hat {\bf k}) \Lambda_{kl} (\hat {\bf k})
\right]
,\label{@projm}\end{eqnarray}
is the
 projection
tensor
to the
physical polarization states,
expressed in terms of
transverse projection matrices
 $\Lambda_{ij} (\hat {\bf k} )= \delta _{ij}-k_i k_j/{\bf k}^2$
of electromagnetism.
It is easy to verify the projection property:
$ \Lambda_{ij,kl}(\hat{{\bf k}})
 \Lambda^{kl}{}_{mn}(\hat{{\bf k}})=
 \Lambda_{ij,mn}(\hat{{\bf k}}).$
As usual,
 $  \beta  \equiv {1}/{k_B T}$, and $k_B$  is the Boltzmann constant.
We have used
the four-vector notation
$x \equiv ({\bf x},t) $  for space
and time. The operator $\hat T$ is the time-ordering operator.

We solve the Schr\"odinger equation  (\ref{2.1})
with the initial condition $\langle {\bf x}|\psi (0) \rangle=\psi({\bf x})$ and
average over the
gravitons. As a result we obtain a density matrix
\begin{eqnarray}
    \rho  ({\bf x}_+ , {\bf x}_-;t) = \langle
 {\bf x}_+ |\psi(t)\rangle \langle \psi(t) | {\bf x}_-\rangle
\label{2.6}\end{eqnarray}
The time evolution of
$ \rho  ({\bf x}_+, {\bf x}_-;t)$
is described by the
evolution equation
\begin{eqnarray}
   \rho  ({\bf x}_{+a} , {\bf x}_{-a} ; t_b) = \int d {\bf x}_{+b}
 d{\bf x}_{-b}
   \,U \left( {\bf x}_{+b}   {\bf x}_{-b}  , t_b | {\bf x}_{+a}
    {\bf x}_{-a} , t_a\right)  \,\rho \left({\bf x}_{+b}
    , {\bf x}_{-b} , t_a\right).
\label{2.6a}\end{eqnarray}
 With the help of the influence functional of Feynman and  Vernon
\cite{FeynmanVernon,FeynmanHibbs,CaldeiraLeggett},
the kernel
of this integral equation
is expressed by the forward--backward  path integral (see
\cite{PI}, and compare with the electromagnetic case
in \cite{HabaKleinert})
\begin{eqnarray}
U \left( {\bf x}_{+b}   {\bf x}_{-b}  , t_b | {\bf x}_{+a}
    {\bf x}_{-a} , t_a\right)  & = &\int  {\cal D}{\bf x}_+
	 {\cal D}{\bf x}_- \exp \left\{ \frac{i}{\hbar }
	 \int^{t_b}_{t_a} dt'\left[ \frac{M}{2} \left[
\dot{\bf x}_+^2    -  \dot{\bf x}_-^2
       \right]
    - V({\bf x}_+) + V ({\bf x}_-)\right] \right\}
\nonumber \\
 &\times & \exp \left[   - \frac{ \epsilon ^2M^2}{2\hbar^2}  \int^{t_b}_{t_a}
dt \int^{t}_{t_a}  dt'
\left(\dot{ x}^k_+
       \dot{ x}_+^l - \dot{ x}^k_- \dot{ x}_-^l\right)
       A_{kl; mn}
\left(\dot{ x}^{'m}_+
       \dot{ x}_+^{'n} - \dot{ x}^{'m}_- \dot{ x}_-^{'n}\right) \right.
 \nonumber \\
 &&  \left.~~\,~~~ -\frac{ \epsilon ^2M^2}{2\hbar^2}
 \int^{t_b}_{t_a}  dt \int^{t}_{t_a}  dt'
\left(\dot{ x}^k_+
       \dot{ x}_+^l - \dot{ x}^k_- \dot{ x}_-^l\right)
		 C_{kl;mn}
       \left(\dot{x}^{'m}_+ \dot{x}^{'n}_+  -
          \dot{x}^{'m}_- \dot{x}^{'n}_-\right)\right]
            \label{2.7}\end{eqnarray}
where $x'$ stands for $x(t')$ and
 $A_{kl;mn}$ and $C_{kl;mn}$ are twice the
real and imaginary parts of (\ref{2.4}).
We are assuming the
wavelengths to be much large than
the size of the object,
which allows us
to neglect the ${\bf x}$-dependence in
$A_{kl;mn}$ and $C_{kl;mn}$ \cite{HH}.
Then we take
the angular average
\begin{equation}
\langle \Lambda_{kl;mn}(\hat{{\bf k}})\rangle
=  \Lambda _{kl;mn}   \equiv  \frac{2}{5}\left[
\frac{1}{2}\left(
\delta _{km}  \delta _{ln}
+\delta _{kn}  \delta _{lm}\right)
    -\frac{1}{3}  \delta _{kl}
   \delta _{mn}\right] ,
\label{@}\end{equation}
and find
\begin{eqnarray}
 C_{kl;mn}&\approx &i\frac{\hbar }{2\pi c^2}
  \Lambda_{kl;mn}\,\partial _t \delta (t-t'),
    \label{2.9}\end{eqnarray}
%
 and at high temperature:
\begin{eqnarray}
A_{kl;mn} =\frac{\hbar }{2\pi c^2}
\frac{1}{ \beta  \hbar  } \Lambda _{kl;mn}
   \, \delta (t-t') ,~~~~~
\label{2.8}\end{eqnarray}
Introducing the traceless dimensionless tensor
\begin{equation}
q^{kl}\equiv\frac{1}{c^2}\left( \dot x^k\dot x^l-\frac{1}{3} \delta^{kl}\dot
x_i\dot x_i\right),
\label{@}\end{equation}
the second exponent can be rewritten as
\begin{eqnarray}
-i\,\frac{ \gamma }2\int _{t_a}^{t_b}dt\,
(q^{kl}_+-q^{kl}_-)
\partial _t
(q^{kl}_++q^{kl}_-)^R
-\frac{ w }{2}\int _{t_a}^{t_b}dt\,
(q^{kl}_+-q^{kl}_-)^2,
\label{@}\end{eqnarray}
where
\begin{equation}
 \gamma \equiv \frac{ \varepsilon ^2 M^2c^2}{10\pi \hbar } ,~~~~
 w \equiv 2 \gamma \frac{k_BT}\hbar .
\label{@}\end{equation}
and the superscript $R$ emphasizes the retardedness of the
$\dot q_{kl}$-terms with respect to the
$ q_{kl}$-terms.
This will be important to fix
the operator order in the
master equation to be derived below.
Note that $ \gamma $ is dimensionless, whereas $w$ is a frequency.

\section{Master Equation for Density Matrix}
Let us derive
an operator
  differential equation for the time evolution of the density matrix.
We denote the Hamilton operator without gravitational field by
\begin{eqnarray}
 \hat H = \frac{\hat {\bf p}^2}{2M} + V ({\bf x}),
\label{4.1}\end{eqnarray}
and proceeding as in the electromagnetic discussion
in Ref.~\cite{HabaKleinert}. The result is the master equation
%
%
%
\begin{eqnarray}
  \partial_t \hat\rho & = &-\frac{i}{\hbar }
{\cal H}\hat  \rho \equiv-\frac{i}{\hbar }
 [\hat H, \hat\rho] -i  \frac{ \gamma  }{2}\left(
\hat q{}^{kl}\hat{\dot  q}{}^{kl} \hat\rho
-\hat\rho\,\hat{\dot  q}{}^{kl}\hat q{}^{kl}
+\hat q{}^{kl} \hat\rho \,\hat{\dot  q}{}^{kl}
-\hat{\dot  q} {}^{kl} \hat\rho \,\hat{ q}{}^{kl} \right)
   - \frac{w}{2}
\left\{[\hat{ q}{}^{kl},[\hat { q}{}^{kl},\hat  \rho ]]
+\frac{\beta ^2\hbar ^2
}{12} [\hat{\dot q}{}^{kl},[\hat {\dot q}{}^{kl},\hat  \rho ]]\right\} .
\label{4.2}\end{eqnarray}
where $  {\hat{\dot x}}{}^k,\, \hat{{\ddot x}}{}^k,\,
\hat{ {\dot q}}{}^{kl},\dots$ are defined
recursively by the commutators
$\hat{\dot { x}}{}^k\equiv {i}[\hat{\cal H},\hat{{x}}^k]/\hbar
 ,
{}~
\hat{\ddot {x}}{}^k\equiv {i}[\hat{\cal H},\hat{\dot{ x}}{}^k]/\hbar,\,
\hat{\dot { q}}{}^{kl}\equiv {i}[\hat{\cal H},\hat{{q}}{}^{kl}]/\hbar
 ,\dots~$.

This equation
will conserve the positivity of the probability
as can be assured by bringing it the the extended
{\em Lindblad form\/} \cite{LB}
\begin{eqnarray}
  \partial_t \hat\rho & = & -\frac{i}{\hbar }
  [\hat H, \hat  \rho ]
-\sum _{n=1}^2  \sum _{kl}
\left(
\frac{1}{2}\hat L_n^{kl} \hat L_n^{kl}{}^\dagger \,\hat \rho
+\frac{1}{2}\hat \rho \,
\hat L_n^{kl}
\hat L_n^{kl}{}^\dagger
- \hat L_n^{kl}{}^\dagger\,\hat \rho\,\hat  L_n^{kl} \right).
\label{3.5Lind}\end{eqnarray}
with the Lindblad operators
\begin{eqnarray}
\hat L_1^{kl}\equiv \frac{ \sqrt{w}}{2 }{\hat q}^{kl},~~~~~~~
\hat L_2^{kl}\equiv \frac{ \sqrt{3w}}{2 }\left( {\hat q}^{kl}
-i\frac{\hbar }{3k_BT}
{\hat{\dot q}}^{kl}
\right).
\label{@}\end{eqnarray}

The master
  equation (\ref{4.2})
 allows us
to calculate
the rate of change of an initial eigenstate
$| i \rangle$ of the Schr\"odinger equation
$H | i \rangle = E_i
 | i \rangle$ by  emission and absorption of
gravitons to lowest order in $G$.
It is a purely classical rate, determined by
the matrix elements
of the $ \gamma $-terms in (\ref{4.2})
 (see \cite{HabaKleinert} for more details):
\begin{equation}
  \partial_t\hat \rho_i \! = \!
-\frac{ \gamma }{2 }
     \left[
\langle i |{\hat q_{kl} } | f \rangle
   \langle f | \hat{\dot q}_{kl}   |i \rangle
\!-\!\langle i |{\hat {\dot q}_{kl} } | f \rangle
   \langle f | \hat{ q}_{kl}   |i \rangle
\! -\!  \langle i |{\hat q_{kl} }| i \rangle
  \langle i |{\hat {\dot q}_{kl} }| i \rangle
 \!+\!
\langle i|\hat{ \dot q}_{kl}   | i \rangle
 \langle i |{\hat q_{kl} }| i \rangle
         \right] =-
 { \gamma }
    \sum_{f \neq i} \omega _{if}
 \langle i |{\hat q_{kl} } | f \rangle
   \langle f |  {\hat q}_{kl}   |i \rangle  .
\label{4.3}\end{equation}
The part of the sum with $f<i$, where the final energy
is lower than the initial energy,
yields the spontaneous decay rate of the system:
\begin{equation}
 \Gamma =2  { \gamma }
    \sum_{f < i} \omega _{if}
 \langle i |{\hat q_{kl} } | f \rangle
   \langle f |  {\hat q}_{kl}   |i \rangle.
\label{@}\end{equation}
For more details of this identification, in particular
the origin of the factor 2,
see Ref.~\cite{HabaKleinert}.
This decay rate
coincides with the perturbative
quantum-mechanical result
obtained from Fermi's golden rule
and the interaction operator $M^{-1}
\int _{t_q}^{t_b}u_{kl} \hat p_k \hat p_l$:
\begin{eqnarray}
   \Gamma  =
\frac{ \varepsilon ^2 }{8\pi^2\hbar } \frac{1  }{c^2M^2}
     \int d \Omega \sum _{h=-2,2}\sum _{f<i}  \omega _{if}
 \left|\epsilon^{kl}(\hat {\bf k},h) \langle f |
 {\hat p}_k {\hat p}_l
    | i \rangle \right|^2
=\frac{ \varepsilon ^2 }{2\pi\hbar } \frac{2  }{5c^2M^2}
     \sum _{f<i}  \omega _{if}
 \left| \langle f |
 {\hat p}_k {\hat p}_l - \frac{1}{3}{{\bf p}^2} \delta _{kl}
    | i \rangle \right|^2,
\label{@qurate}\end{eqnarray}
In terms of $ \Gamma $, the right-hand side of (\ref{4.3}) can be written as
\begin{eqnarray}
  \partial_t\hat \rho_i  = - \Gamma
+ { \gamma }
    \sum_{f < i} \omega _{if}
 \langle i |{\hat q_{kl} } | f \rangle
   \langle f |  {\hat q}_{kl}   |i \rangle
- { \gamma }
    \sum_{f > i} \omega _{if}
 \langle i |{\hat q_{kl} } | f \rangle
   \langle f |  {\hat q}_{kl}   |i \rangle  .
\label{4.}\end{eqnarray}
By spontaneous emission, $\hat \rho_i$ decays like
$e^{- \Gamma t}$.
The rest is due to induced emissions and absorptions
contributing to
 the
rate of change of $\hat \rho_i$.

The total
 level width is increased further
 by the induced
 emissions and absorptions
 proportional to $w$
 in (\ref{4.2}):
\begin{eqnarray}
 \partial_t \rho_i & = &  -  {w}
\left\{  \sum _f
 \langle i |\hat q_{kl} | f \rangle
 \langle f |\hat q_{kl} | i \rangle
 -
 \langle i |\hat q^2_{kl} | i \rangle ^2
  +\frac{ \beta ^2\hbar ^2}{12}
 \left[\sum _f
 \langle i |\hat{\dot  q}_{kl} | f \rangle
 \langle f |\hat{\dot  q}_{kl} | i \rangle
-
\langle i |\hat{\dot q}^2_{kl} | i \rangle ^2
 \right]\right\}  .
\label{4.5}\end{eqnarray}
These are due to
further induced emission and absorption processes
(see Ref.~\cite{HabaKleinert} for details).

Only the last term in this expressions is
caused by the
quantum nature
of the emitted gravitons.
In fact, Eq.~(\ref{@qurate})
agrees with the classical result for the
emitted power  by inserting an energy per graviton
$\hbar  \omega _{if}$ into the sum
and replacing $ \omega _{if}^2$ times  the square of the matrix elements
by
the classical expression
$(M^{-2}\partial _t q_{kl})^2/2$.
Recall that according to the correspondence principle,
we find
the quantum-mechanical result for the transition $|i\rangle $ to $|f\rangle$
from the classical formula by replacing
$q_{ij}(t)\rightarrow \langle f|\hat q_{ij}|i\rangle e^{-i \omega _{if}t}+$c.c.
$=2|\langle f|\hat q_{ij}|i\rangle | \cos(\omega _{if}t- \delta )$
from which we
deduces the
replacement of the classical $(\partial _t q_{ij})^2$ by the temporal
average of the quantum-mechanical
$ 2 \omega _{if}^2  |\langle f|\hat q_{ij}|i\rangle |^2$.

A possible future application of our master equation
(\ref{3.5Lind})
could be to neutrino oscillation experiments, modifying the
discussions of Refs.~\cite{Adler,Lisi}.
\section{Langevin equation}
At high and moderately high temperatures,
the fluctuations
of ${\bf y}$ are small due to the last term in the exponent
of (\ref{2.7}). We therefore
 introduce average and difference variables
of the forward and backward paths in (\ref{2.7}):
\begin{eqnarray}
 {\bf x} \equiv   \left({\bf x}_+ + {\bf x}_-\right)/2,~~~~~
  {\bf y}  =  {\bf x}_+ - {\bf x}_-,
\label{@}\end{eqnarray}
and
expand $V({\bf x}_\pm)=
V({\bf x}\pm {\bf y}/2)$ in ${\bf y}$ up to the third order.
Then  the exponent in Eq.~(\ref{2.7})
becomes
\begin{eqnarray}
 &&\!\!\!\!\!\!\!\!\!\!\!\!\!
 \frac{i}{\hbar} \int^{t_b}_{t_a}\!\!dt \left\{
M\dot {\bf y} \dot {\bf x}
 -{\bf y}\BF\nabla V ({\bf x})
 -\frac{1}{24}
{y}_i
{y}_j
y_k
\nabla_i\nabla_j\nabla_k V ({\bf x})
-2 \frac{\gamma \hbar}{c^4}
  \left[
  \dot{\bf x}^2 (\ddot{\bf x} \dot{\bf y})
 +\frac{1}{3} (\dot{\bf y} \dot{\bf x}) (\ddot{\bf x}\dot{\bf x})
    +\frac{1}{3} (\ddot{\bf y} \dot{\bf y}) (\dot{\bf x}\dot{\bf y})
+\frac{1}{4}  \dot{\bf y}^2 (\ddot{\bf y}\dot{\bf x})
 \right]
\right.\nonumber \\ & & \left.  ~~~~~~~~
+2i\,\frac{w\hbar }
{c^4}
  \left[  \dot{\bf y}^2 \dot{\bf x}^2+\frac{1}{3}
    ( \dot{\bf y} \dot{\bf x})^2\right]
  \right\},
\label{3.1}\end{eqnarray}
where
fourth and
higher  orders in ${\bf y}(t)$ are neglected.
Neglecting also the quadratic terms ${\bf y}(t)$, extremization
with respect of ${\bf y}(t)$ gives the classical equation
of motion with radiation damping
\begin{eqnarray}
   M \ddot {\bf x} +{\BF \nabla} V -2 \frac{ \gamma\hbar }{c^4}
 \frac{d}{dt} \left[  \ddot {\bf x}\, \dot{\bf x}^2
  +\frac{1}{3}  \dot {\bf x}\, (\ddot{\bf x}\dot{\bf x})
  \right] = 0
\label{3.2}\end{eqnarray}
%
{}From this equation we can calculate
the energy dissipation of the gravitational body
due to the coupling to the bath of gravitons:
\begin{eqnarray}
  \frac{d}{dt} \left\{ \left[\frac{M \dot{\bf x}^2}{2} + V
 ({\bf x})\right]
 - \frac{2}{3}
 \frac{ \gamma\hbar }{c^4}
    \frac{d}{dt} ( \dot{\bf x}^2)^2\right\}
 =  -2
 \frac{ \gamma\hbar }{c^4}
 \left[ \ddot {\bf x}^2 \dot{\bf x}^2
      +\frac{1}{3} (\ddot{\bf x}\dot{\bf x})^2\right].
\label{3.3}\end{eqnarray}
Expressed in terms of $q_{kl}$,
the right-hand side yields a rate of energy loss
\begin{equation}
\dot E=-\gamma \hbar \,\dot q_{kl}^2=\frac{4G}{5} \,\dot q_{kl}^2
{}.
\label{@}\end{equation}
This is the formula
derived for
classical gravitational radiation.
The
extra term in the brackets is the gravitational
analog of Schott's term in the
energy balance of electromagnetic radiation damping discussed in
\cite{HabaKleinert,Rohr1,Rohr}.

Interpreting this equation  twice over $t$ we can conclude
that the time-average of the energy decreases owing to the
dissipative term on the r.h.s.~of (\ref{3.3}).
The quadratic fluctuations of ${\bf y}(t)$ in the path integral
(\ref{2.9}) are governed by the quadratic exponent (\ref{3.1}),
which can be written as a
path integral
\begin{eqnarray}
 && \int {\cal D}{\bf x} {\cal D}{\bf y}
\exp\left\{
 i \int^{t_b}_{t_a}dt
\left[-\frac{{\cal M}}{2}
   \dot y_k  \Omega _{kl} \dot y_l
- \frac{i}{\hbar}
    y_k \dot u_k\right]
\right\}
F({\bf x})
\label{3.4}\end{eqnarray}
where
$\dot{\bf u}(t)$ abbreviates the left-hand side of Eq.~(\ref{3.2}),
\begin{equation}
{\cal M}\equiv \frac{4w}{3c^4},~~~~ \Omega _{kl}\equiv
\dot x_k
\dot x_l+
 3\dot {\bf x}^2
 \delta _{kl}.
\label{@}\end{equation}
The exponent can be completed quadratically to
\begin{eqnarray}
\int^{t_b}_{t_a}dt
\left\{
 - \frac{{\cal M}}{2}
  \left[\left(\dot y_k - \frac{i}{{\cal M}\hbar}
 \Omega ^{-1}_{kl} u_l\right)
   \Omega _{kr}
\left(\dot y_r - \frac{i}{{\cal M}\hbar}\Omega ^{-1} _{rl} u_l
\right)
\right]
  -\frac{1}{2{\cal M}\hbar^2}  u_l \Omega ^{-1}
_{lr} u_r \right\}.
\label{3.4a}\end{eqnarray}
We now introduce
the stochastic variable
\begin{eqnarray}
\BF\eta \equiv   \Omega ^{-1/2} {\bf u} ,
\label{@}\end{eqnarray}
which has the  white-noise
correlation function
\begin{equation}
\langle  \eta _i(t) \eta _j(t')\rangle={\cal M}\hbar ^2 \delta _{ij} \delta
(t-t')
=\frac{4w\hbar ^2}{3c^4}
  \delta _{ij} \delta (t-t')
{}.
\label{@}\end{equation}
Ignoring the
Jacobian of the transformation ${\bf x}(t) \rightarrow {\bf u}(t)$,
the path integral is seen to be equivalent
 to the Langevin equation
\begin{eqnarray}
 M \ddot {\bf x} +\BF \nabla V({\bf x}) -
{2} \frac{\gamma \hbar }{c^4}\frac{d}{dt} \left[ \ddot {\bf x} \dot {\bf x}^2
 +\frac{1}{3}  \dot {\bf x}(\ddot{\bf x} \dot{\bf x})^2\right]
 = \frac{d}{dt} \Omega ^{1/2}\BF \eta.
\label{3.5}\end{eqnarray}
Note that this Langevin equation
is purely classical, since the constants
$ \gamma \hbar $ and $w \hbar^2 $
are $\hbar$-independent.
In electromagnetism, a similar equation has been derived in
Ref.~\cite{HabaKleinert}
solved in Ref.~\cite{HabaKleinert2}.

The omitted Jacobian consists of a product of two
Jacobians for the transformations
${\bf x}\rightarrow \dot {\bf u}$ and
$\dot {\bf u}\rightarrow \BF \eta$.
For the latter, it is equal to Det$\, \partial _t \Omega ^{1/2}$
which  corresponds to an effective action $\log$\,Det$\, \partial _t \Omega
^{1/2}$ proportional to $ \delta (0)$,
This vanishes in dimensional regularization \cite{PI}.
The other Jacobian is nontrivial
and awaits a proper evaluation along the lines of
Ref.~\cite{KL}, and the result may modify Eq.~(\ref{3.5}).

The equation is applicable only for high temperatures.
If we want to apply it to moderately high temperatures,
quantum effects appear.
To lowest order in $ \hbar$, these arise
from the first nonleading Taylor expansion term of
$\coth [ \beta  \hbar \Omega_{\sbf k}/2]$ in (\ref{2.4}).
This term changes the $ \delta $-function in
$A_{kl;mn}$
of
 Eq. (\ref{2.8})
to
\begin{eqnarray}
  \delta (t-t')\rightarrow
  \delta (t-t') - \frac{\hbar^2\beta ^2}{24}
    \ddot \delta (t-t').
\label{@extrat}\end{eqnarray}
This leads to a change
 $\BF \eta \rightarrow  \BF \eta
-{\hbar^2\beta ^2}
 \ddot{\BF \eta}/24$
on the right-hand side of the Langevin equation (\ref{3.5}).

Let us point out that our derivation
of the classical equation
(\ref{3.3})
in the present quantum mechanical context
is much simpler and more direct than
the purely classical derivations
in the literature.
There one always start from extended objects
to find the forces exerted upon a body by
its own  gravitational
radiation. The subsequent limit of a point particle
is quite delicate.
For reviews on the subject see
Refs.~\cite{Rohr1,Mino,Wald}

{}~\\
This research is supported by a grant from a governmental
 German university
 support program
HSP III - Potsdam.

\end{document}